\documentclass[final,authoryear,5p,times,twocolumn]{elsarticle}
\usepackage{graphicx}

\usepackage{amssymb}

\journal{Radiation Measurements}

\begin{document}

\begin{frontmatter}



\title{Vacuum ultraviolet 5d$^1$4f$^9$-4f$^{10}$ emission of Ho$^{3+}$ ions in alkaline-earth fluorides}


\author{E A Radzhabov$^{1,2}$}
\ead{eradzh@igc.irk.ru}

\author{V. Nagirnyi$^{3}$ }

\author{A I Nepomnyashchikh$^{1,2}$}

\address{$^1$Vinogradov Institute of Geochemistry, Russian Academy of Sciences, Favorskii street 1a, P.O.Box 4019, 664033 Irkutsk, Russia}

\address{$^2$Irkutsk State University, Physics department, Gagarin boulevard 20, 664003 Irkutsk, Russia}

\address{$^3$Institute of Physics, University of Tartu, Riia 142, 51014 Tartu, Estonia}

\begin{abstract}
Time-resolved emission, excitation as well as emission decay curves of CaF$_2$, SrF$_2$, BaF$_2$ doped with HoF$_3$ were investigated. 
Most intensive emission bands near 168 nm, having long decay time, belong to spin-forbidden transitions from 5d$^1$4f$^9$ high spin (HS) states to ground $^5$I$_8$ states of Ho$^{3+}$ ions. Weak spin allowed 5d$^1$4f$^9$(LS)-4f$^{10}$ emission band at 158 nm was observed only in CaF$_2$-Ho crystals. Spin allowed and spin-forbidden excitation bands were observed in all crystals near 166 and 155 nm respectively.  Fast component of spin-forbidden emissions due to multiphonon relaxation to low-lying 4f$^{10}$ Ho$^{3+}$ level  was observed for all crystals.

\end{abstract}


\end{frontmatter}


\section{Introduction}
Spectroscopy of f-d transitions of rare-earth ions doped into crystals remains relatively less investigated field of optical spectroscopy \citep{Makhov2001}. The spectroscopic investigation of 4f$^n$-5d$^1$4f$^{n-1}$ transitions of rare-earth ions (hereafter simply 4f-5d)  in the vacuum ultraviolet region could help in search of new scintillators or new ultraviolet lasers materials.

For holmium two weak emissions in the vacuum ultraviolet region of the spectrum were observed at 158 nm in YF$_3$ and at 167 nm in LiYF$_4$ excited at 130 nm \citep{Peijzel2002}. These emissions were attributed to Ho$^{3+}$ spin-forbidden transitions from 4f$^9$5d  to the ground state \citep{Peijzel2002}. The spectrum of the LiYF$_4$:Ho$^{3+}$ crystal under F$_2$ laser pumping (at 157 nm) shows several bands in the spectral range from 157.6 to 190 nm. The fluorescence peaks were assigned to the spin-allowed and the spin-forbidden transitions between the levels of the 4f$^9$5d and the levels of the 4f$^{10}$ electronic configuration of the Ho$^{3+}$ ion \citep{Sarantopoulou1999}.  Ho$^{3+}$ vacuum ultraviolet emission of alkaline-earth fluorides using continuous discharge lamp excitation were shortly presented in our previous papers \citep{Radzhabov2011, Radzhabov2012}.

The main goal of this paper is the time-resolved spectroscopy of Ho$^{3+}$ ultraviolet and vacuum ultraviolet emission in CaF$_2$, SrF$_2$, BaF$_2$ with concentration of HoF$_3$ dopants ranging from 0.01 to 0.3 molar \%.

\section{Experimental}

Crystals were grown in vacuum in a graphite crucible by the Stockbarger method \citep{Radzhabov2012}.  Graphite crucible contains three cylindrical cavities 10 mm in diameter and 80 mm long, which allows to grow three crystals of {\O}10x50 mm in dimensions with different impurity concentrations at the same time. Typically samples {\O}10 mm x 1mm cutted from grown roads were used for optical measurements.

Time-resolved spectra and decay curves were recorded using synchrotron radiation at the SUPERLUMI station of HASYLAB at DESY (Hamburg, Germany) as described in paper \citep{Kirm2001}. Measurements were performed in the short time window (1-5 ns), in the long time window (55-80 ns), and in the time-integrated regime.

Additionally absorption and excitation spectra were measured using grating monochromator VM4 (by LOMO) under excitation of deuterium lamp L7293-50 with MgF$_2$ window (by Hamamatsu) and a solar-blind photomultiplier FEU142 as light detector.

\section{Results}

\begin{figure}[h]
\includegraphics[width=22pc]{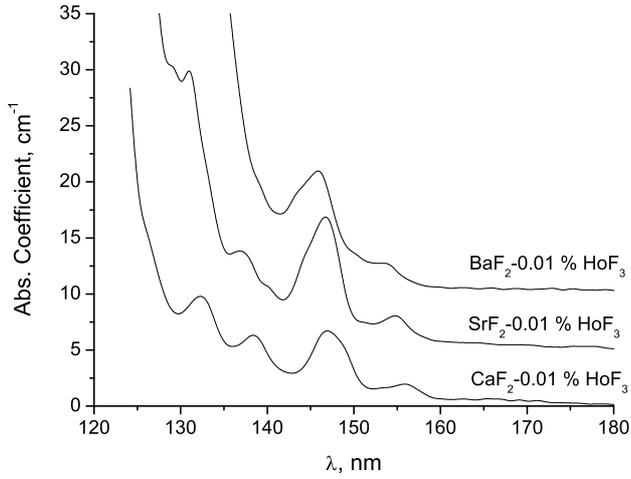}
\caption{\label{abs} Absorption spectra of Ho$^{3+}$ 5d-4f transitions in CaF$_2$, SrF$_2$, BaF$_2$ crystals doped with 0.01 molar \% of HoF$_3$ at 295K. Spectra were separated by 5 cm$^{-1}$ for better viewing.}
\end{figure}

Absorption spectra of Ho$^{3+}$ ion in all three crystals show several bands below 160 nm (Fig.\ref{abs}). All bands are shifted to less wavelength by few nm in a row of CaF$_2$ to BaF$_2$. Shift of absorption  4f-5d and emission 5d-4f bands in a row of CaF$_2$ to BaF$_2$ are known for other rare-earth  ions, it is due to decrease of crystal field splitting of 5d levels \citep{Radzhabov2008}. 

\begin{figure}[h]
\includegraphics[width=21pc]{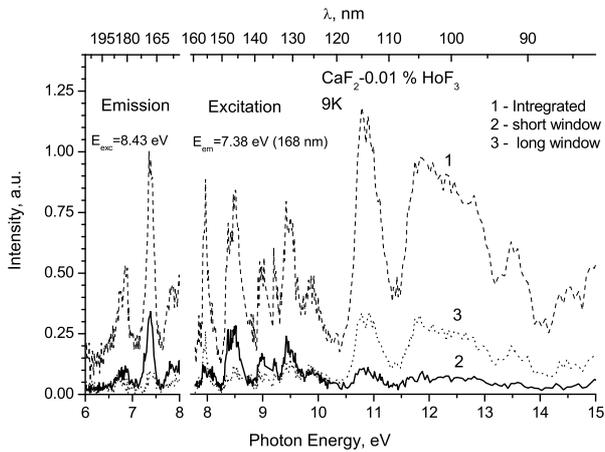}
\caption{\label{CaF-exc} Emission and excitation spectra of Ho$^{3+}$ 5d-4f transitions in CaF$_2$ crystals doped with 0.01 molar \% of HoF$_3$ at 9K.}
\end{figure}

\begin{figure}[h]
\includegraphics[width=21pc]{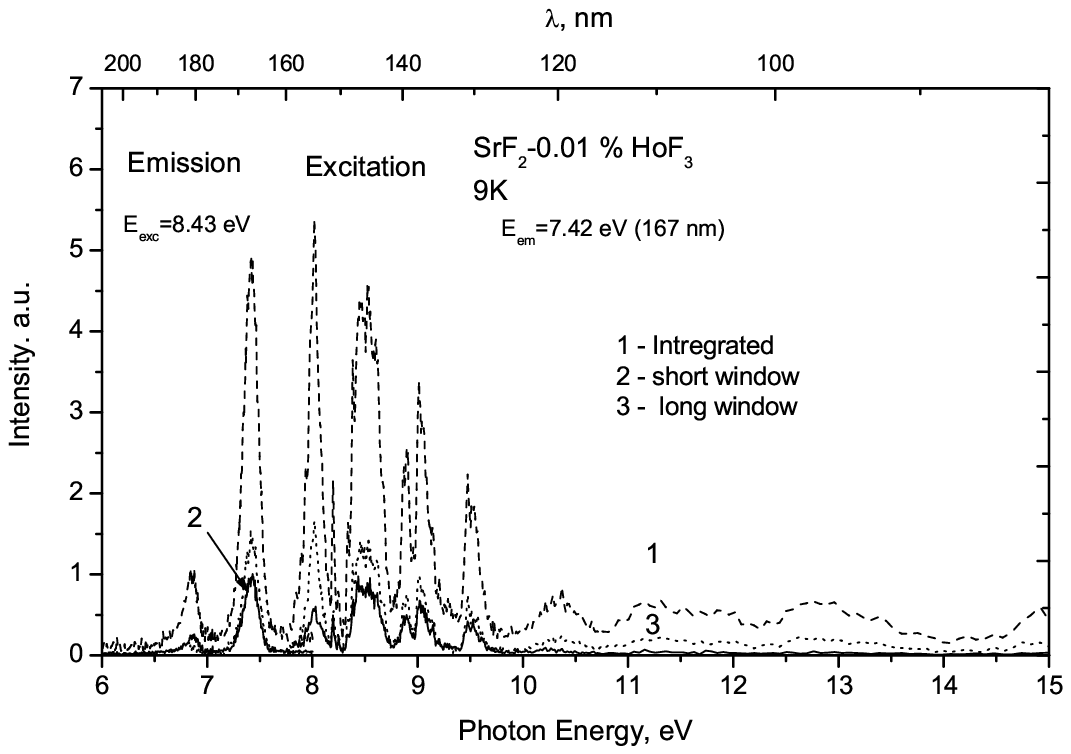}
\caption{\label{SrF-exc} Emission and excitation spectra of Ho$^{3+}$ 5d-4f transitions in SrF$_2$ crystals doped with 0.01 molar \% of HoF$_3$ at 9K.}
\end{figure}

\begin{figure}[h]
\includegraphics[width=21pc]{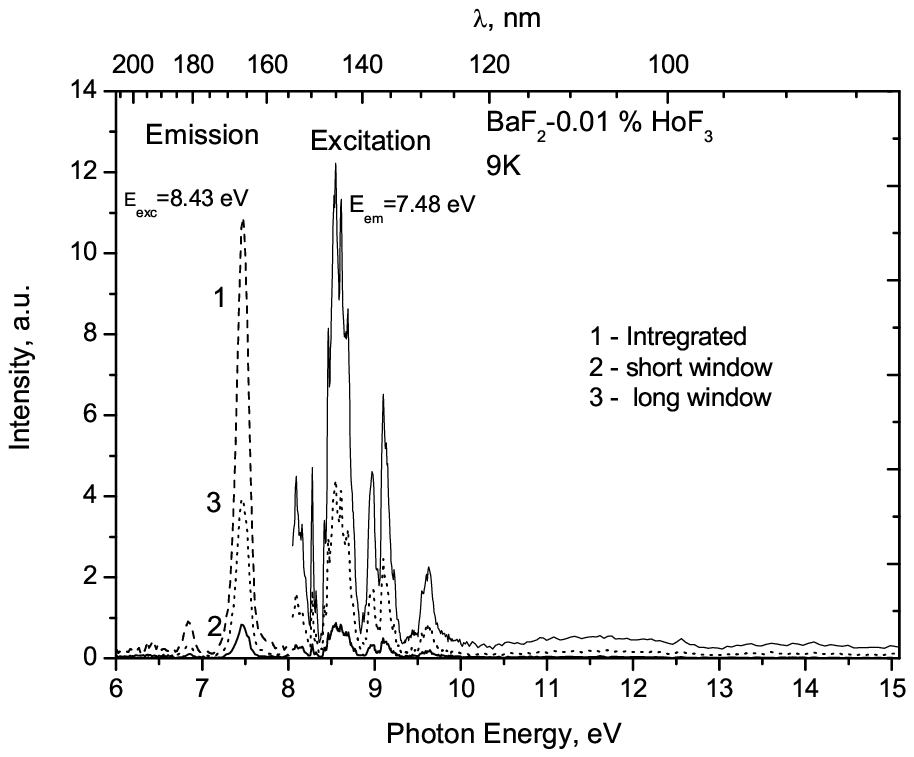}
\caption{\label{BaF-exc} Emission and excitation spectra of Ho$^{3+}$ 5d-4f transitions in BaF$_2$ crystals doped with 0.01 molar \% of HoF$_3$ at 9K.}
\end{figure}

\begin{figure}[h]
\includegraphics[width=21pc]{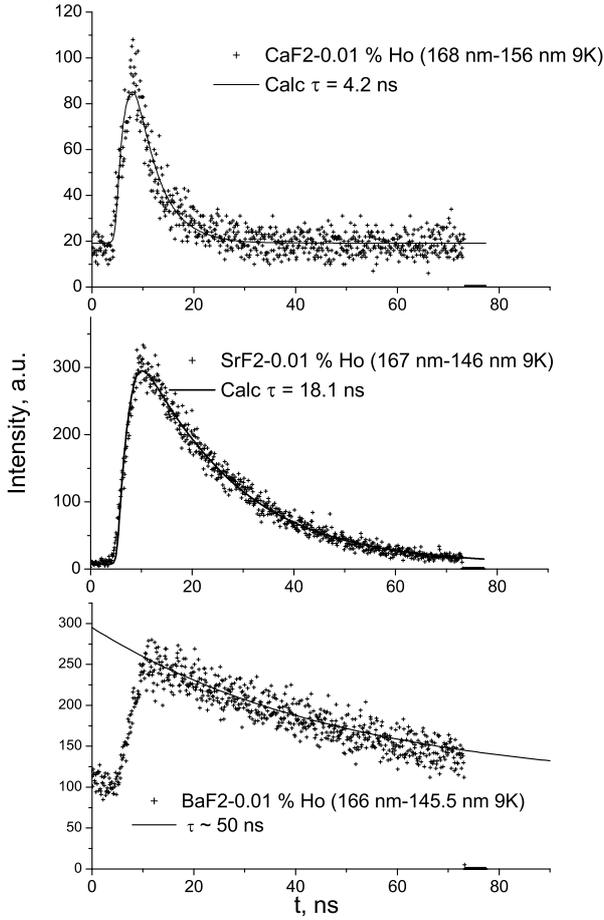}
\caption{\label{decay} Decay of Ho$^{3+}$ emission bands near 167nm in BaF$_2$, SrF$_2$, CaF$_2$ crystals doped with 0.01 molar \% of HoF$_3$ at 9K.}
\end{figure}

Two 5d-4f emission bands at 167-169 and 180-183 nm were observed with intensity ratio of near 10:1 (Fig.\ref{CaF-exc}-\ref{BaF-exc}). The bands were assigned to the transitions from 5d lowest energy level to $^5$I$_8$ and $^5$I$_7$ levels of 4f shell. Four main excitation bands 155, 146, 138 and 130 nm correlate with four absorption bands (compare Figs.\ref{abs} - \ref{BaF-exc}). Both emission bands contain slow and fast decay components. The decay time of fast component increased by an order of magnitude in a row of CaF$_2$ to BaF$_2$ (Fig.\ref{decay}).  The decay times of fast components under illumination into lowest energy excitation band near 155 nm were equal to ~50, 15.7 and 4.2 ns in BaF$_2$, SrF$_2$, and CaF$_2$ respectively (Fig.\ref{decay}). The decay times become few ns longer under excitation into the 146 nm band.

The 5d-4f emission bands are relatively weak in CaF$_2$ while they become more intensive in SrF$_2$ and BaF$_2$. Apart to this 4f-4f emission bands drastically decreased from CaF$_2$ to SrF$_2$, BaF$_2$. Two 5d-4f Ho$^{3+}$ bands in BaF$_2$, SrF$_2$  evidently belong to transition to $^5$I$_8$ ground state and to $^5$I$_7$ first excited state of 4f$^{10}$ configuration. Emission spectrum in CaF$_2$  is a combination of 5d(LS)-4f and 5d(HS)-4f bands. 

\section{Discussion}

\begin{figure}[h]
\includegraphics[width=20pc]{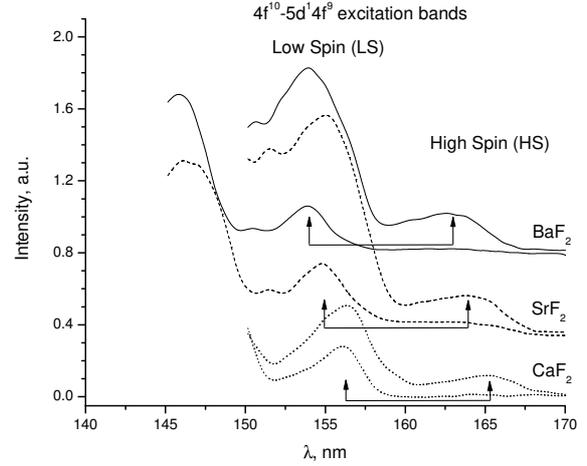}
\caption{\label{Ho_exc} Excitation spectra of Ho$^{3+}$ 5d-4f transitions in CaF$_2$, SrF$_2$, BaF$_2$ crystals doped with 0.01 molar \% of HoF$_3$ (lower curves) and  by 0.03 molar \% of HoF$_3$ at 295K. Spectra were separated for best viewing. Splitting of first excitation bands by transitions to low spin and high spin 5d4f9 states shown by arrows.}
\end{figure}

The zero-phonon Ho$^{3+}$ 4f-5d absorption and excitation lines were observed in CaF$_2$ at 157.1 nm \citep{Szszurek1985} and 157.5 nm \citep{Pieterson2002}  respectively. No spin-forbidden transitions were observed in excitation spectrum, possibly due to the low Ho$^{3+}$ concentration \citep{Pieterson2002}. To find the spin-forbidden bands we have measured excitation spectra for integrated emission of all f-f bands within 220-800 nm wavelength range of all three hosts with concentration of HoF$_3$ from 0.01 to 0.3 molar \% (Fig \ref{Ho_exc}). Spectra were measured at room temperature because f-f emission under f-d excitation substantially decreased with decreasing temperature. New weak excitation bands were found in all hosts at long wavelength side of first 4f-4f5d bands with separation near 9 nm (see Fig.\ref{Ho_exc}). Bands were questionable in samples with 0.01 \% of Ho, but were evident in samples containing 0.3 \% of HoF$_3$ (see Fig \ref{Ho_exc}). These excitation bands are attributed to transitions to high-spin 5d4f states. Energy gaps between Ho$^{3+}$ 4f5d(HS) and 4f5d(LS) states is near 3500 cm$^{-1}$ for all CaF$_2$, SrF$_2$, BaF$_2$ hosts, which is close to that observed in LiYF$_4$-Ho (3460cm$^{-1}$ ) \citep{Pieterson2001}. We conclude that the emission band near 158 nm in CaF$_2$ -Ho belongs to spin-allowed transitions from 5d4f(HS) and the most intensive emission band near 169 nm belongs to spin-forbidden transitions from 5d4f(LS) to ground state. Energy splitting between 4f5d(HS) and 4f5d(LS) emission and excitation bands in CaF$_2$ -Ho is the same (compare Fig.\ref{CaF-exc} and Fig.\ref{Ho_exc}). No spin-allowed transitions were observed in SrF$_2$ -Ho or BaF$_2$ -Ho crystals (see also \citep{Radzhabov2012}). The absence of spin-allowed emission bands in SrF$_2$ and BaF$_2$ is not yet clear.

It is known that nonradiative transfer probability from some f level to underlying levels exponentially decrease with energy gap between levels (so called energy gap law) \citep{Henderson1989, Sole2005}. Reformulation of this law in term of number of effective phonons allows to conclude that nonradiative processes are dominant for processes involving less than 4-6 effective phonons \citep{Henderson1989}. Comparison of ultraviolet Nd emission in a number orthoborate and orthophosphate crystals lead to the conclusion that nonradiative transition from 5d level to lower lying 4f level is predominant when energy gap is less than the energy of 5 phonons \citep{Wegh2001}. 

The frequencies of a longitudinal optic phonon obtained from optical data at 5 K are 484, 397 and 346 cm$^{-1}$ for CaF$_2$, SrF$_2$, BaF$_2$ respectively \citep{Hayes1974}. The highest observed Ho$^{3+}$ 4f$^{10}$ level is $^3$K$_7$, transitions to this level were observed at 170.5 nm in LiYF$_4$ \citep{Peijzel2002}. 4f$^9$5d$^1$ (HS) excitation bands were measured at 162.5, 164, 165 nm in CaF$_2$, SrF$_2$, BaF$_2$, respectively (see Fig.\ref{Ho_exc}). Therefore the energy gap between these two levels are near 2000, 2300, 2900 cm$^{-1}$ or around 4, 6, 8 phonon frequencies in CaF$_2$, SrF$_2$, BaF$_2$, respectively. It follows that nonradiative multiphonon energy transfer of Ho$^{3+}$ from 5d4f$^{9}$(HS) to 4f$^{10}$ ($^3$K$_7$) level should be most effective in CaF$_2$ and much less effective in BaF$_2$. As the result, the 5d4f$^{9}$-4f$^{10}$ band intensity is increased and 4f$^{10}$-4f$^{10}$ bands intensities sharply decreased in a row of CaF$_2$ to BaF$_2$. The fast decay components of spin-forbidden bands obviously should be assigned to multiphonon transition from 5d to  4f$^{10}$. According to increase the energy gap the fast decay time become drastically longer in a row of CaF$_2$, SrF$_2$, BaF$_2$ (see Fig.\ref{decay}). The nonradiative relaxation rate decreased with decreasing temperature, as number of phonons becomes lesser \citep{Henderson1989}. According to this the Ho$^{3+}$ f-f line intensities are several times decreased from 78K to 6.7K under f-d excitation \citep{Radzhabov2012}. The host and temperature variations of 5d-4f and 4f-4f intensity and decay times after 4f-5d excitation could be described in term of nonradiative multiphonon energy transfer from high spin lowest 5d level to the low-lying 4f$^{10}$($^3$K$_7$) level. 

\begin{figure}[h]
\includegraphics[width=20pc]{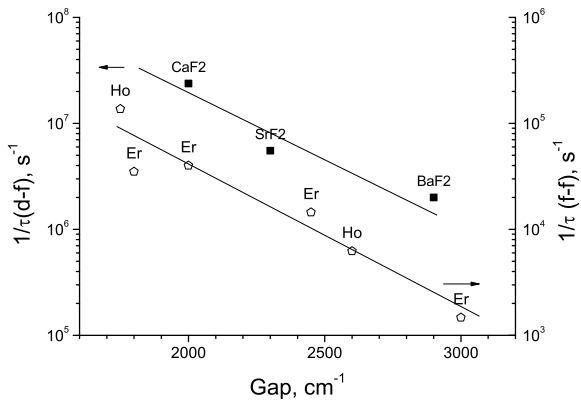}
\caption{\label{Gap}Energy-gap dependence of the multiphonon transition rate for f-f transition in SrF$_2$-RE taken from \citealp{Riseberg1968} and for d-f transition of Ho$^{3+}$ in alkaline-earth fluorides}
\end{figure}

The dependence of the multiphonon relaxation probability on the energy gap, disregarding any selection rules, is of the form  \citep{Dijk1983}

\begin{center}
$k_{NR}=\beta e^{-\alpha \Delta E}$
\end{center}

where $\beta$ and $\alpha$ are constants which are characteristic of the particular crystal and $\Delta$E is the electronic energy "gap" between the 4f levels. However, the $\beta$ varies by a factor of 10$^5$.  This wide variation is due to a considerable dependence of $\beta$ on the vibrational coupling to the host lattice \citep{Dijk1983}. While there are number of papers on measurements of multiphonon f-f relaxation, but to our knowledge, no such measurements on d-f relaxation have been made.

We plot the inverse decay times of fast components of spin-forbidden 5d$^1$4f$^9$-$^5$I$_8$ bands near 168 nm versus the energy gap between 5d and nearest 4f levels (Fig.\ref{Gap}). Three hosts CaF$_2$, SrF$_2$, BaF$_2$ have the same crystal structure and slightly differ with anion-cation distances. Therefore, in the first approximation, the data for d-f multiphonon relaxation in three alkaline earth fluorides could be compared with that of f-f relaxation in SrF$_2$, which we took from paper \citep{Riseberg1968}. Note that the slope $\alpha$ for both dependencies is nearly the same while the constant $\beta$ is thee order higher for d-f multiphonon relaxation (see Fig.\ref{Gap}). The constant $\beta$  reflects the interaction of orbitals with phonons. Therefore it is not surprising that constant $\beta$ is much larger for d-f relaxation than for f-f multiphonon relaxation, because the d-orbital has much greater intersection with orbital of surrounding host ions.

\section{Conclusion}

Ho$^{3+}$ 5d$^1$4f$^9$-4f$^{10}$ emission in vacuum ultraviolet  were found in CaF$_2$, SrF$_2$, BaF$_2$ crystals. Two bands are observed : strongest ones at  166-168 nm  and weaker bands at 182-184 nm. The bands are associated with spin forbidden transition from 5d$^1$4f$^9$ to $^5$I$_8$, $^5$I$_7$ respectively. Weaker spin-allowed 5d$^1$4f$^9$-$^5$I$_8$ band at 158 nm is observed in CaF$_2$ only. 
Decrease of total intensity of f-f bands in compare with that of d-f bands as well as increase fast decay times of spin-forbidden emission in a row of CaF$_2$-BaF$_2$ caused by increasing  number of phonons in the process of nonradiative multiphonon energy transfer from lowest 5d level to nearest 4f$^{10}$ excited level. 

\section*{Acknowledgments}
The authors gratefully acknowledge V. Kozlovskii for growing the crystals investigated in this work.  The work was partially supported by Federal Target Program ”Scientiﬁc and scientiﬁc-pedagogical personnel of innovative Russia” in 2009-2013 (contracts 8367, 8382).

\label{}





\bibliographystyle{model2-names}
\bibliography{Holmium}







\end{document}